\documentclass[12pt,prd,tightenlines,nofootinbib,showpacs,showkeys]{revtex4}
\newcommand{\be}{\begin{equation}}
\newcommand{\ee}{\end{equation}}
\usepackage{bm}
\usepackage{graphics}
\usepackage{rotating}
\usepackage{epsfig}
\begin{document}
\title{\begin{flushright}{\rm\normalsize SSU-HEP-09/9\\[5mm]}\end{flushright}
Relativistic effects in the double S- and P-wave \\
charmonium production in $e^+e^-$ annihilation}
\author{E.N. Elekina}
\author{A.P. Martynenko}
\affiliation{Samara State University, Pavlov Street 1, Samara
443011, Russia}

\begin{abstract}
On the basis of perturbative QCD and the relativistic quark model we
calculate relativistic and bound state corrections in the production
processes of a pair of S-wave and P-wave charmonium states.
Relativistic factors in the production amplitude connected with the
relative motion of heavy quarks and the transformation law of the
bound state wave function to the reference frame of the moving S-
and P-wave mesons are taken into account. For the gluon and quark
propagators entering the production vertex function we use a
truncated expansion in the ratio of the relative quark momenta to
the center-of-mass energy $\sqrt{s}$ up to the second order. The
exact relativistic treatment of the wave functions makes all such
second order terms convergent, thus allowing the reliable
calculation of their contributions to the production cross section.
Relativistic corrections to the quark bound state wave functions in
the rest frame are considered by means of the Breit-like potential.
It turns out that the examined effects change essentially the
nonrelativistic results of the cross section for the reaction
$e^++e^-\to J/\Psi(\eta_c)+\chi_{cJ}(h_c)$ at the center-of-mass
energy $\sqrt{s}=10.6$ GeV.
\end{abstract}

\pacs{13.66.Bc, 12.39.Ki, 12.38.Bx}

\keywords{Hadron production in $e^+e^-$ interactions, Relativistic quark model}

\maketitle

\section{Introduction}

The large rate for the exclusive double charmonium production
measured at the Belle and BaBar experiments \cite{Belle,BaBar}
reveals definite problems in the theoretical description of these
processes \cite{BL1,Chao,Qiao}. Many theoretical efforts were made
in order to improve the calculation of the production cross section
$e^++e^-\to J/\Psi+\eta_c$. They included the analysis of other
production mechanisms for the state $J/\Psi+\eta_c$ \cite{BLB,BGL}
and the calculation of different corrections which could change
essentially the initial nonrelativistic result
\cite{Chao1,Ma,BC,BLL,ZGC,Bodwin2,EM2006,Ji,He,AVB,Bodwin4}. Despite
the evident successes achieved on the basis of NRQCD, the light cone
method, quark potential models in order to resolve the discrepancy
between the theory and experiment, the double charmonium production
in $e^+e^-$ annihilation remains an interesting task. On the one
hand, the reason is that there exist the production processes of the
P- and D-wave charmonium states which should be investigated as the
production of S-wave states. On the other hand, the variety of the
used approaches and the model parameters in this problem raises the
question about the comparison of the obtained results resulting in a
better understanding of the quark-gluon dynamics. Two sources of the
enhancement of the nonrelativistic cross section for the double
charmonium production are revealed to the present: the radiative
corrections of order $O(\alpha_s)$ and relative motion of c-quarks
forming the bound states.

In this work we continue the investigation of the exclusive double
charmonium production in $e^+e^-$ annihilation on the basis of a
relativistic quark model \cite{EM2006,EFGM2009,apm2005,rqm5} in the
case of S- and P-wave charmonium states. The relativistic quark
model provides the solution in many tasks of heavy quark physics. In
particular, it gives the possibility to study the question about a
broadening of the meson wave functions due to the account of special
corrections which can lead to the increase of the double charmonium
production cross sections. Thus, the aim of this study consists in
the calculation of the relativistic effects in the processes
$e^++e^-\to J/\Psi(\eta_c)+\chi_{cJ}(h_c)$ on the basis of a
relativistic approach to the quarkonium production suggested in
Refs.\cite{EM2006,EFGM2009}.

\section{General formalism}

We consider the following reactions $e^++e^-\to
J/\Psi(\eta_c)+\chi_{cJ}(h_c)$, where the final state consists of
the pair of S-wave ($J/\Psi$ or $\eta_c$) and P-wave ($\chi_{c0}$,
$\chi_{c1}$, $\chi_{c2}$ or $h_c$) charm mesons. The diagrams that
give contributions to the amplitude of these processes in the
leading order of the QCD coupling constant $\alpha_s$ are presented
in Fig.1. Two other diagrams can be obtained by corresponding
permutations. There are two stages of the production process. In the
first stage, which is described by perturbative QCD, the virtual
photon $\gamma^\ast$ produces four heavy c-quarks and $\bar
c$-antiquarks with the following four-momenta:

\begin{figure}
\centering
\includegraphics[width=5.cm]{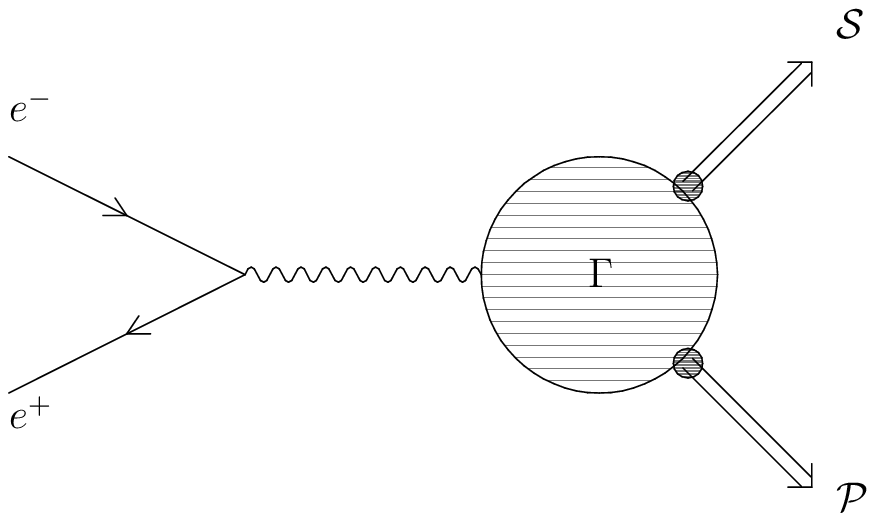}\hspace*{0.4cm}
\includegraphics[width=11cm]{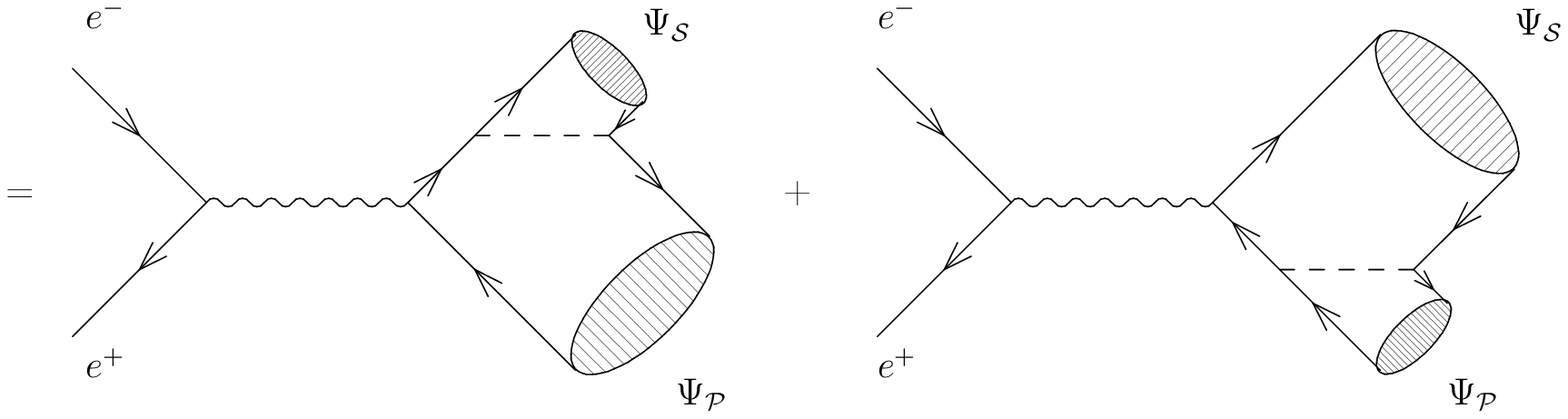}
\caption{The production amplitude of a pair of S- and P-wave
charmonium states in $e^+e^-$ annihilation. ${\cal S}$ denotes the
S-state meson and ${\cal P}$ the P-wave meson. The wavy line shows
the virtual photon and the dashed line corresponds to the gluon.
$\Gamma$ is the production vertex function.}
\end{figure}

\begin{equation}
p_{1,2}=\frac{1}{2}P\pm p,~~(p\cdot P)=0;~~q_{1,2}=\frac{1}{2}Q\pm q,~~(q\cdot Q)=0,
\end{equation}
where $P(Q)$ are the total four-momenta, $p=L_P(0,{\bf p})$,
$q=L_P(0,{\bf q})$ are the relative four-momenta obtained from the
rest frame four-momenta $(0,{\bf p})$ and $(0,{\bf q})$ by the
Lorentz transformation to the system moving with the momenta $P,Q$.
In the second nonperturbative stage, quark-antiquark pairs form the
final mesons.

Let consider the production amplitude of the $S$-wave vector state
($J/\Psi$) and $P$-wave states $\chi_{cJ}$ (J=0,1,2), which can be
presented in the form \cite{F1973,BP,rqm5,EM2006}:
\begin{equation}
{\cal M}(p_-,p_+,P,Q)=\frac{8\pi^2\alpha\alpha_{s}{\cal
Q}_c}{3s}\bar v(p_+)\gamma^\beta u(p_-)\int\frac{d{\bf
p}}{(2\pi)^3}\int\frac{d{\bf q}}{(2\pi)^3}\times
\end{equation}
\begin{displaymath}
\times Sp\left\{\Psi^{\cal
S}(p,P)\Gamma_1^{\beta\nu}(p,q,P,Q)\Psi^{\cal
P}(q,Q)\gamma_\nu+\Psi^{\cal
P}(q,Q)\Gamma_2^{\beta\nu}(p,q,P,Q)\Psi^{\cal S}(p,P)\gamma_\nu
\right\},
\end{displaymath}
where $\alpha_{s}$ is the QCD coupling constant, $\alpha$ is the
fine structure constant and ${\cal Q}_c$ is the $c$ quark electric
charge. The relativistic S- and P-wave functions of the bound quarks
$\Psi^{\cal S, \cal P}$ accounting for the transformation from the
rest frame to the moving one with four momenta $P,Q$, are
\begin{eqnarray}
\Psi^{\cal S}(p,P)&=&\frac{\Psi_0^{\cal S}({\bf
p})}{\left[\frac{\epsilon(p)}{m}
\frac{(\epsilon(p)+m)}{2m}\right]}\left[\frac{\hat v_1-1}{2}+\hat
v_1\frac{{\bf p}^2}{2m(\epsilon(p)+ m)}-\frac{\hat{p}}{2m}\right]\cr
&&\times\hat{\varepsilon_S}^\ast(1+\hat v_1) \left[\frac{\hat
v_1+1}{2}+\hat v_1\frac{{\bf p}^2}{2m(\epsilon(p)+
m)}+\frac{\hat{p}}{2m}\right],
\end{eqnarray}
\begin{eqnarray}
\Psi^{\cal P}(q,Q)&=&\frac{\Psi_0^{\cal P}({\bf q})}
{\left[\frac{\epsilon(q)}{m}\frac{(\epsilon(q)+m)}{2m}\right]}
\left[\frac{\hat v_2-1}{2}+\hat v_2\frac{{\bf q}^2}{2m(\epsilon(q)+
m)}+\frac{\hat{q}}{2m}\right]\cr &&\times\hat{\varepsilon}_{\cal
P}^\ast(Q,S_z)(1+\hat v_2) \left[\frac{\hat v_2+1}{2}+\hat
v_2\frac{{\bf q}^2}{2m(\epsilon(q)+ m)}-\frac{\hat{q}}{2m}\right],
\end{eqnarray}
where $v_1=P/M_{\cal S}$, $v_2=Q/M_{\cal P}$; ${\varepsilon_S}$ is
the polarization vector of the vector charmonium $J/\Psi$;
$\varepsilon_{\cal P}(Q,S_z)$ is the polarization vector of the
spin-triplet state $\chi_{cJ}$, $\epsilon(p)=\sqrt{p^2+m^2}$ and $m$
is the $c$ quark mass. At the leading order in $\alpha_s$ the vertex
functions $\Gamma_{1,2}^{\beta\nu}(p,P;q,Q)$ can be written as
\begin{equation}
\Gamma_1^{\beta\nu}(p,P;q,Q)= \gamma_\mu\frac{(\hat l-\hat
q_1+m)}{(l-q_1)^2-m^2+i\epsilon} \gamma_\beta D^{\mu\nu}(k_1)+
\gamma_\beta\frac{(\hat p_1-\hat l+m)}{(l-p_1)^2-m^2+i\epsilon}
\gamma_\mu D^{\mu\nu}(k_1),
\end{equation}
\begin{equation}
\Gamma_2^{\beta\nu}(p,P;q,Q)=\gamma_\beta\frac{(\hat q_2-\hat
l+m)}{(l-q_2)^2-m^2+i\epsilon} \gamma_\mu D^{\mu\nu}(k_2)+
\gamma_\mu\frac{(\hat l-\hat p_2+m)}{(l-p_2)^2-m^2+i\epsilon}
\gamma_\beta D^{\mu\nu}(k_2),
\end{equation}
where the gluon momenta are $k_1=p_1+q_1$, $k_2=p_2+q_2$ and
$l^2=s=(P+Q)^2=(p_-+p_+)^2$, $p_-$, $p_+$ are four momenta of the
electron and positron. The dependence on the relative momenta of
$c$-quarks is presented both in the gluon propagator $D_{\mu\nu}(k)$
and quark propagator as well as in the relativistic wave functions
(3), (4). Taking into account that the ratio of the relative quark
momenta $p$ and $q$ to the energy $\sqrt{s}$ is small, we expand the
inverse denominators of quark and gluon propagators as follows:
\begin{equation}
\frac{1}{(l-q_{1,2})^2-m^2}=\frac{2}{s}\left[1-\frac{2M_{\cal
S}^2-M_{\cal P}^2-4m^2}{2s}-\frac{2q^2}{s}\pm\frac{4(lq)}
{s}+\frac{16(lq)^2}{s^2}+\cdots\right],
\end{equation}
\begin{equation}
\frac{1}{(l-p_{1,2})^2-m^2}=\frac{2}{s}\left[1-\frac{2M_{\cal
P}^2-M_{\cal S}^2-4m^2}{2s}-\frac{2p^2}{s}\pm\frac{4(lp)}
{s}+\frac{16(lp)^2}{s^2}+\cdots\right],
\end{equation}
\begin{equation}
\frac{1}{k_{2,1}^2}=\frac{4}{s}\left[1-\frac{4(p^2+q^2+2pq)}{s}\pm\frac{4(lp+lq)}{s}+
\frac{16}{s^2}[(lp)^2+(lq)^2+2(lp)(lq)]+\cdots\right].
\end{equation}
In the expansions (7)-(9) we accounted for terms of second order in
relative momenta $p$ and third order in relative momenta $q$.
Substituting (7)-(9), (3)-(4) in (2) we preserve relativistic
factors entering the denominators of the relativistic wave functions
(3)-(4), but in the numerator of the amplitude (2) we take into
account corrections of second order in $|{\bf p}|/m$ and up to
fourth order in $|{\bf q}|/m$. This provides the convergence of the
resulting momentum integrals. Then the angular integrals are
calculated using the following relations:
\begin{equation}
\int p_\mu p_\nu \Psi_0^{\cal S}({\bf p})\frac{d{\bf
p}}{(2\pi)^3}=-\frac{\sqrt{4\pi}}{3}(g_{\mu\nu}-v_{1~\mu}v_{1~\nu})\int_0^\infty
p^4R_{\cal S}(p)dp,
\end{equation}
\begin{equation}
\int q_\mu\Psi_0^{\cal P}({\bf q})\frac{d{\bf
q}}{(2\pi)^3}=-i\varepsilon_{{\cal
P}~\mu}(Q,L_z)\frac{1}{\pi\sqrt{6}}\int_0^\infty q^3 R_{\cal
P}(q)dq,
\end{equation}
\begin{equation}
\int q_\alpha q_\beta q_\gamma \Psi_0^{\cal P}({\bf q})\frac{d{\bf
q}}{(2\pi)^3}=\frac{i}{5\pi\sqrt{6}}[\varepsilon_\gamma(Q,L_z)P_{\alpha\beta}+
\varepsilon_\alpha(Q,L_z)P_{\gamma\beta}+\varepsilon_\beta(Q,L_z)
P_{\alpha\gamma}]\int_0^\infty q^5R_{\cal P}(q)dq,
\end{equation}
where $P_{\alpha\beta}=(g_{\alpha\beta}-v_{2~\alpha}v_{2~\beta})$,
$R_{\cal S}(p)$, $R_{\cal P}(q)$ are the radial momentum wave
functions of S- and P-wave charmonium states,
$\varepsilon_\mu(Q,L_z)$ is the polarization vector in orbital
space. The integrals in (10) and (12) look formally divergent, but
the original momentum integrals contain also definite relativistic
factors which lead to their convergence. For a specific P-wave
state, summing over $S_z$ and $L_z$ in the amplitude (2) can be
further simplified as \cite{Kuhn}
\begin{equation}
\sum_{S_z,L_z}<1,L_z;1,S_z|J,J_z>\varepsilon^\ast_{{\cal
P}~\alpha}(Q,L_z)\varepsilon^\ast_{{\cal P}~\beta}(Q,S_z)=
\cases{\frac{1}{\sqrt{3}}(g_{\alpha\beta}-v_{2~\alpha}v_{2~\beta}),~~~J=0,\cr
\frac{i}{\sqrt{2}}\epsilon_{\alpha\beta\sigma\rho}v_2^\sigma\epsilon^{\ast~\rho}(Q,J_z),~~~J=1,\cr
\epsilon_{\alpha\beta}(Q,J_z),~~~J=2,\cr}
\end{equation}
where $<1,L_z;1,S_z|J,J_z>$ are the Clebsch-Gordon coefficients.
Calculating the trace in the amplitude (2) by means of expressions
(3)-(6), (13) and the system FORM \cite{FORM}, we find that the
tensor parts of the four amplitudes describing the production of S-
and P-wave charmonium states in the used approximation have the
following structure:
\begin{equation}
S_{1,\beta}(J/\Psi+\chi_{c0})=A_1\varepsilon^\ast_{S~\beta}+A_2v_{1~\beta}(v_2\varepsilon^\ast_S)+
A_3v_{2~\beta}(v_2\varepsilon^\ast_S),
\end{equation}
\begin{equation}
S_{2,\beta}(J/\Psi+\chi_{c1})=B_1\varepsilon_{\alpha\lambda\gamma\beta}v_1^\alpha
v_2^\lambda\varepsilon^{\ast~\gamma}(Q,J_z)(v_2\varepsilon^\ast_S)+
B_2\varepsilon_{\alpha\lambda\gamma\beta}v_2^\alpha
\varepsilon_S^{\ast~\lambda}\varepsilon^{\ast~\gamma}(Q,J_z)+
\end{equation}
\begin{displaymath}
+B_3v_{1~\beta}\varepsilon_{\alpha\lambda\gamma\sigma}v_1^\alpha
v_2^\lambda\varepsilon_S^{\ast~\gamma}\varepsilon^{\ast~\sigma}(Q,J_z)+
B_4v_{2~\beta}\varepsilon_{\alpha\lambda\gamma\sigma}v_1^\alpha
v_2^\lambda\varepsilon_S^{\ast~\gamma}\varepsilon^{\ast~\sigma}(Q,J_z),
\end{displaymath}
\begin{equation}
S_{3,\beta}(J/\Psi+\chi_{c2})=\varepsilon^\ast_{\alpha\sigma}(Q,J_z)[C_1\varepsilon^{\ast~\alpha}_S
g_{\sigma\beta}+C_2v_1^\alpha(v_2\varepsilon^\ast_S)g_{\beta\sigma}+C_3
v_1^\alpha\varepsilon^{\ast~\sigma}_Sv_{1~\beta}+C_4v_1^\alpha\varepsilon^{\ast~\sigma}_Sv_{2~\beta}+
\end{equation}
\begin{displaymath}
+C_5v_1^\alpha v_1^\sigma \varepsilon^\ast_{S~\beta}+C_6v_1^\alpha
v_1^\sigma v_1^\beta(_2\varepsilon^\ast_S)+C_7v_1^\alpha v_1^\sigma
v_{2~\beta}(v_2\varepsilon^\ast_S)],
\end{displaymath}
\begin{equation}
S_{4,\beta}(\eta_c+h_c)=D_1v_{1~\beta}(v_1\varepsilon^\ast(Q,L_z))+D_2v_{2~\beta}
(v_1\varepsilon^\ast(Q,L_z))+D_3\varepsilon^\ast_\beta(Q,L_z),
\end{equation}
where the coefficients $A_i$, $B_i$, $C_i$, $D_i$ can be presented
as sums of terms containing the factors $u=M_{\cal P}/(M_{\cal
P}+M_{\cal S})$, $\kappa=m/(M_{\cal P}+M_{\cal S})$ and
$C_{ij}=c^i(p)c^j(q)=[(m-\varepsilon(p))/(m+\varepsilon(p))]^i
[(m-\varepsilon(q))/(m+\varepsilon(q))]^j$, preserving terms with
$i+j\leq 2$, and $r^2=(M_{\cal P}+M_{\cal S})^2/s$ up to terms of
order $O(r^4)$. Exact analytical expressions for these coefficients
are sufficiently lengthy (compare with the results written in
Appendix A of our previous paper \cite{EFGM2009}), so, we present in
Appendix A of this work only their approximate numerical form using
the observed meson masses and the c-quark mass $m=1.55$ GeV.

Introducing the scattering angle $\theta$ between the electron
momentum ${\bf p}_e$ and the momentum ${\bf P}$ of the $J/\Psi$
meson, we can calculate the differential cross section
$d\sigma/d\cos\theta$ and then the total cross section $\sigma$ as a
function of $r^2$. We find it useful to present the charmonium
production cross sections in the following form ($k=0,1,2,3$
corresponds to $\chi_{c0}$, $\chi_{c1}$, $\chi_{c2}$ and $h_c$):

\begin{equation}
\sigma(J/\Psi(\eta_c)+\chi_{cJ}(h_c))=\frac{\alpha^2\alpha_s^2{\cal
Q}_c^2\pi
r^2\sqrt{1-r^2}\sqrt{1-r^2(2u-1))}}{6912\kappa^2u^9(1-u)^9}\frac{|\tilde
R_{\cal S}(0)|^2|\tilde R'_{\cal P}(0)|^2}{s(M_{\cal P}+M_{\cal
S})^8}\sum_{i=0}^7F_i^{(k)}(r^2)\omega_i,
\end{equation}
where the functions $F_i^{(k)}$ (k=0,1,2,3) are written explicitly
in Appendix B,
\begin{equation}
\tilde R_{\cal S}(0)=\frac{1}{2\pi^2}\int_0^\infty
p^2R_S(p)\frac{(\epsilon(p)+m)}{2\epsilon(p)}dp,
\end{equation}
\begin{equation}
\tilde R'_{\cal P}(0)=\frac{1}{3}\sqrt{\frac{2}{\pi}}\int_0^\infty
q^3R_P(q)\frac{(\epsilon(q)+m)}{2\epsilon(q)}dq.
\end{equation}
The parameters $\omega_i$ can be expressed in terms of momentum
integrals $I_n$, $J_n$ as follows:
\begin{equation}
I_n=\int_0^\infty p^2R_{\cal
S}(p)\frac{(\epsilon(p)+m)}{2\epsilon(p)}\left(\frac{m-\epsilon(p)}
{m+\epsilon(p)}\right)^ndp, J_n=\int_0^\infty q^3R_{\cal
P}(q)\frac{(\epsilon(q)+m)}{2\epsilon(q)}\left(\frac{m-\epsilon(q)}
{m+\epsilon(q)}\right)^ndq,
\end{equation}
\begin{equation}
\omega_0=1,~~ \omega_1=\frac{I_1}{I_0},~~
\omega_2=\frac{I_2}{I_0},~~ \omega_3=\omega_1^2,~~
\omega_4=\frac{J_1}{J_0},~~ \omega_5=\frac{J_2}{J_0},~~
\omega_6=\omega_4^2, ~~\omega_7=\omega_1\omega_4.
\end{equation}

On the one side, in the potential quark model the relativistic
corrections, connected with the relative motion of heavy c-quarks,
enter the production amplitude (2) and the cross section (18)
through the different relativistic factors. They are determined in
the final expression (18) by the specific parameters $\omega_i$. The
momentum integrals which determine the parameters $\omega_i$ are
convergent and we calculate them numerically, using the wave
functions obtained by the numerical solution of the Schr\"odinger
equation. The exact form of the wave functions $\Psi_0^{\cal S}({\bf
p})$ and $\Psi_0^{\cal P}({\bf q})$ is important for improving the
accuracy of the calculation of the relativistic effects. It is
sufficient to note that the double charmonium production cross
section $\sigma(s)$ in the nonrelativistic approximation contains
the factor $|R_{\cal S}(0)|^2 |R'_{\cal P}(0)|^2$. Small changes of
the numerical values of the bound state wave functions at the origin
lead to substantial changes of the final results. In the approach
based on nonrelativistic QCD this problem is closely related to the
determination of the color-singlet matrix elements for the
charmonium \cite{BBL}. Thus, on the other side, there are
relativistic corrections to the bound state wave functions
$\Psi_0^{\cal S}({\bf p})$, $\Psi_0^{\cal P}({\bf q})$. In order to
take them into account, we suppose that the dynamics of a $c\bar
c$-pair is determined by the QCD generalization of the standard
Breit Hamiltonian \cite{pot1,pot2,pot3}:
\begin{equation}
H=H_0+\Delta U_1+\Delta U_2,~~~H_0=2\sqrt{{\bf
p}^2+m^2}-2m-\frac{C_F\alpha_s}{r}+Ar+B,
\end{equation}
\begin{equation}
\Delta U_1(r)=-\frac{C_F\alpha_s^2}{4\pi r}\left[2\beta_0\ln(\mu
r)+a_1+2\gamma_E\beta_0
\right],~~a_1=\frac{31}{3}-\frac{10}{9}n_f,~~\beta_0=11-\frac{2}{3}n_f,
\end{equation}
\begin{equation}
\Delta U_2(r)=-\frac{C_F\alpha_s}{2m^2r}\left[{\bf p}^2+\frac{{\bf
r}({\bf r}{\bf p}){\bf p}}{r^2}\right]+\frac{\pi
C_F\alpha_s}{m^2}\delta({\bf r})+\frac{3C_F\alpha_s}{2m^2r^3}({\bf
S}{\bf L})-
\end{equation}
\begin{displaymath}
-\frac{C_F\alpha_s}{2m^2}\left[\frac{{\bf S}^2}{r^3}-3\frac{({\bf
S}{\bf r})^2}{r^5}-\frac{4\pi}{3}(2{\bf S}^2-3)\delta({\bf
r})\right]-\frac{C_AC_F\alpha_s^2}{2mr^2},
\end{displaymath}
where $n_f$ is the number of flavors, $C_A=3$ and $C_F=4/3$ are the
color factors of the SU(3) color group. For the dependence of the
QCD coupling constant $\alpha_s(\mu^2)$ on the renormalization point
$\mu^2$ we use the leading order result
\begin{equation}
\alpha_s(\mu^2)=\frac{4\pi}{\beta_0\ln(\mu^2/\Lambda^2)}.
\end{equation}
The typical momentum transfer scale in a quarkonium is of order of
the quark mass, so we set the renormalization scale $\mu=m$ and
$\Lambda=0.168$ GeV, which gives $\alpha_s=0.314$ for the charmonium
states. The parameters of the linear potential $A=0.18~GeV^2$ and
$B=-0.16$ GeV have usual values of quark models. Starting with the
Hamiltonian (23) we construct the effective potential model based on
the Schr\"odinger equation and find its numerical solutions in the
case of S- and P-wave charmonium \cite{FFS}. The details of the used
model are presented in Appendix C. Then we calculate the matrix
elements entering in the expressions for the parameters $\omega_i$
and obtain the value of the production cross sections at
$\sqrt{s}$=10.6 GeV. Basic parameters which determine our numerical
results are collected in Table I. The comparison of the obtained
results with the previous calculations \cite{BL1,Chao,BLL1,Chao2008}
and experimental data \cite{Belle,BaBar} is presented in Table II.

\begin{table}
\caption{Numerical values of the relativistic parameters (19), (20),
(22) in the double charmonium production cross section (18).}
\bigskip
\begin{ruledtabular}
\begin{tabular}{|c|c|c|c|c|c|c|}
Meson  $(c\bar c)$& $n^{2S+1}L_J$ & $J^{PC}$ &$\tilde R_{\cal
S}(0)$, $GeV^{3/2}$ &$\tilde R'_{\cal P}(0)$, $GeV^{5/2}$ &
$\omega_1({\cal S})$ or $\omega_4({\cal P})$& $\omega_2({\cal S})$ or $\omega_5({\cal P})$\\
\hline $J/\Psi$& $1^3S_1$&$1^{--}$ & 0.81 &
--- & -0.20  & 0.0078
\\  \hline
$\eta_c$&$1^1S_0$&$0^{-+}$  & 0.92 & --- & -0.20  & 0.0087
\\  \hline
$\chi_{c0}$&$1^3P_0$&$0^{++}$  & --- & 0.19 & -0.15  & 0.0065
\\  \hline
$\chi_{c1}$&$1^3P_1$&$1^{++}$   & --- & 0.18 & -0.14  & 0.0065
\\  \hline
$\chi_{c2}$&$1^3P_2$&$2^{++}$   & --- & 0.18 & -0.15  & 0.0065
\\  \hline
$h_c$&$1^1P_1$&$1^{+-}$   & --- & 0.18 & -0.14  & 0.0065
\\  \hline
\end{tabular}
\end{ruledtabular}
\end{table}

\section{Numerical results and discussion}

In this paper we have investigated the role of relativistic effects
in the production processes of S- and P-wave mesons $(c\bar c)$ in
the quark model. In the present study of the production amplitude
(2) we kept relativistic corrections of two types. The first type is
determined by several functions depending on the relative quark
momenta  ${\bf p}$ and ${\bf q}$ arising from the gluon propagator,
the quark propagator and the relativistic meson wave functions. The
second type of corrections originates from the perturbative
treatment of the quark-antiquark interaction operator which leads to
the different wave functions $\Psi_0^{\cal S}({\bf p})$ and
$\Psi_0^{\cal P}({\bf q})$ for the S-wave and P-wave charmonium
states, respectively. In addition, we systematically accounted for
the bound state corrections working with the observed masses of
S-wave mesons ($J/\Psi$, $\eta_c$) and P-wave mesons ($\chi_{cJ}$,
$h_c$). The calculated masses of S-wave and P-wave charmonium states
agree well with experimental values \cite{PDG} (see Table III). Note
that the basic parameters of the model are kept fixed from the
previous calculations of the meson mass spectra and decay widths
\cite{rqm5,rqm1,QWG}. The strong coupling constant entering the
production amplitude (2) is taken to be $\alpha_s$=0.24 in
accordance with the relation (26) at $\mu=2m$.

Numerical results and their comparison with several previous
calculations and experimental data are presented in Table II.
Theoretically, there were two studies of the production
$J/\Psi(\eta_c)+\chi_{cJ}(h_c)$ in $e^+e^-$ annihilation in NRQCD
\cite{BL1,Chao}.  They give 2.4 fb and 6.7 fb for the production of
$J/\Psi+\chi_{c0}$. Such spread in results is explained by the
different numerical values of the used parameters, i.e. the matrix
elements, the mass of c-quark $m$ and the strong coupling constant
$\alpha_s$. The third investigation of the process
$J/\Psi+\chi_{c0}$ was done in the light-front formalism in
\cite{BLL1} where the result 14.4 fb was obtained. The essential
growth of the cross section in \cite{BLL1} at $\sqrt{s}=10.6$ GeV is
connected with the use of specific light cone wave functions
describing the relative motion of heavy c-quarks. The fourth study
of the reaction $e^++e^-\to J/\Psi+\chi_{c0}$ was devoted to the
next-to-leading order QCD corrections \cite{Chao2008}. Here it was
shown that a sharp increase of the production cross section
($\sigma=17.9$ fb) can be derived with the account of NLO in
$\alpha_s$ contributions.

The exclusive double charmonium production cross section presented
in the form (18) is convenient for a comparison with the results of
NRQCD. Indeed, in the nonrelativistic limit, when $u=1/2$,
$\kappa=1/4$, $\omega_i=0$ ($i\geq 1)$), $r^2=16m^2/s$, the cross
section (18) coincides with the calculation in \cite{BL1}. In this
limit the functions $F_0^{(k)}(r^2)$ transform into corresponding
functions $F_k$ from \cite{BL1}. When we take into account bound
state corrections working with observed meson masses, we get
$u=M_{\cal P}/(M_{\cal P}+M_{\cal S}) \not = 1/2$,
$\kappa=m/(M_{\cal P}+M_{\cal S})\not =1/4$. This leads to the
modification of the general factor in (18) in comparison with the
nonrelativistic theory and the form of the functions $F_0^{(k)}$
(see \cite{BL1}). It follows from the numerical values of the
parameters $\omega_i$, presented in Table I, that the relativistic
corrections amount to $15\div 20 \%$ in the production amplitude.
Moreover, the relativistic effects decrease the values of the
parameters $R_{\cal S}(0)$, $R'_{\cal P}(0)$, which transform into
$\tilde R_{\cal S}(0)$, $\tilde R'_{\cal P}(0)$. In all considered
reactions $e^++e^-\to J/\Psi(\eta_c)+\chi_{cJ}(h_c)$ the
relativistic effects increase the nonrelativistic cross section, but
in the case of the production $\eta_c+h_c$ the sum of bound state
plus relativistic corrections decreases the nonrelativistic cross
section. It is necessary to point out once again that the essential
effect on the value of the production cross sections
$J/\Psi(\eta_c)+\chi_{cJ}(h_c)$ belongs to the parameters $\tilde
R_{\cal S}(0)$,  $\tilde R'_{\cal P}(0)$, $\alpha_s$, $m$. Small
changes in their values can lead to significant changes for the
production cross sections. Comparing the values of the parameters
$R_{\cal S}(0)$ ($|R_{J/\Psi,\eta_c}(0)|^2=0.9;1.2~GeV^{3/2}$),
$R'_{\cal P}(0)$ ($|R'_{\cal P}(0)|^2=0.043~GeV^{5/2}$) obtained in
this study on the basis of quark model and in \cite{Chao2008} we see
that the values of the radial wave functions at the origin are very
close, but our value for the derivative of the radial wave function
at the origin is slightly smaller. The calculation of radiative
corrections $O(\alpha_s)$ to the nonrelativistic cross section of
the production $J/\Psi+\chi_{c0}$ was done recently in
\cite{Chao2008}. It evidently shows that one-loop corrections are
considerable (factor $K=2.8$ to nonrelativistic result). As a
result, the total value of the cross section $J/\Psi+\chi_{c0}$
significantly increases. It should be noted that the difference
between the theory and both BaBar and Belle experiments became
threatening.

We presented a systematic treatment of relativistic effects in the
S- and P-wave double charmonium production in $e^+e^-$ annihilation.
We explicitly separated two different types of relativistic
contributions to the production amplitudes. The first type includes
the relativistic $v/c$ corrections to the wave functions and their
relativistic transformations. The second type includes the
relativistic $p/\sqrt{s}$ corrections emerging from the expansion of
the quark and gluon propagators. The latter corrections were taken
into account up to the second order. It is important to note that
the expansion parameter $p/\sqrt{s}$ is very small. In our analysis
of the production amplitudes we correctly take into account
relativistic contributions of order $O(v^2/c^2)$ for the S-wave
meson and corrections of orders $O(v^2/c^2)$ and $O(v^4/c^4)$ for
the P-wave mesons. We cannot keep corrections of order $O(v^4/c^4)$
for the S-wave part of the amplitude (2) because they become
divergent if we use expansions (7)-(9). Therefore the basic
theoretical uncertainty of our calculation is connected with the
omitted terms of order $O({\bf p}^4/m^4)$. Taking into account that
the average value of the heavy quark velocity squared in the
charmonium is $<v^2>=0.3$, we expect that they should not exceed
30\% of the obtained relativistic contribution. These theoretical
errors in the calculated production cross section at $\sqrt{s}=10.6$
GeV are shown directly in Table II. We have neglected the terms in
the cross section (18) containing the product of $I_n$ and $J_n$
with summary index $\ge 2$ because their contribution has been found
negligibly small. There are no another comparable uncertainties
related to the choice of $m$ or any other parameters of the model,
since their values were fixed from our previous consideration of
meson and baryon properties \cite{rqm1,rqm5}.

\begin{table}
\caption{Comparison of the obtained results with previous
theoretical predictions and experimental data.}
\bigskip
\begin{ruledtabular}
\begin{tabular}{|c|c|c|c|c|c|c|c|}
State  & $\sigma_{BaBar}\times$ & $\sigma_{Belle}\times $
&$\sigma_{NRQCD}$& $\sigma$ $(fb)$ &$\sigma$ $(fb)$ & $\sigma$
$(fb)$ & Our result \\
$H_1H_2$   &$ Br_{H_2\to charged\ge 2}$ & $Br_{H_2\to charged\ge 2}$
&$(fb)$ \cite{BL1}& \cite{Chao}  &\cite{BLL1}& \cite{Chao2008}  &  $(fb)$  \\
 &$(fb)$ \cite{BaBar} & $(fb)$ \cite{Belle}    &  &  &    &    &
\\   \hline
$J/\Psi+\chi_{c0}$ & $10.3\pm 2.5^{+1.4}_{-1.8}$ & $6.4\pm
1.7\pm 1.0$ &  $2.40\pm 1.02$& 6.7& 14.4  & 17.9(6.35)  &  $4.79\pm 0.80$ \\
\hline  $J/\Psi+\chi_{c1}$ &  & & $0.38\pm 0.12$& 1.1 &  & &
$1.07\pm 0.23$ \\  \hline $J/\Psi+\chi_{c2}$ &  & &  $0.69\pm 0.13$ &1.6 & & & $1.10\pm 0.13$ \\
\hline $\eta_c+h_c$ & &  & $0.308\pm 0.017$&  & &  &  $0.24\pm 0.02$ \\  %\hline
\end{tabular}
\end{ruledtabular}
\end{table}

\acknowledgments

The authors are grateful to D. Ebert, R.N. Faustov and V.O. Galkin
for useful comments and discussions. The work is performed under the
financial support of the Federal Program "Scientific and pedagogical
personnel of innovative Russia"(grant No. NK-20P/1).

\appendix

\section{The coefficients $A_i$, $B_i$, $C_i$, $D_i$ entering in
the production amplitudes (14)-(17)}

These coefficients are the sums of the terms containing the
parameters $u=M_{\cal P}/(M_{\cal P}+M_{\cal S})$ and
$\kappa=m/(M_{\cal P}+M_{\cal S})$. We present $A_i$, $B_i$, $C_i$,
$D_i$ in numerical form using the observed meson masses and the mass
of c-quark $m=1.55$ GeV.\\[1mm]

\vspace{1mm}
{\underline {$e^++e^-\to J/\Psi+\chi_{c0}$}}

\begin{equation}
A_1=-7.05+\frac{18.55}{r^2}+0.013r^2+C_{20}(7.05-\frac{18.55}{r^2}+0.013r^2)+
\end{equation}
\begin{displaymath}
+C_{02}(5.42-\frac{5.41}{r^2}-0.013r^2)+C_{10}(68.16-\frac{63.92}{r^2}-12.15r^2+0.089r^4)+
\end{displaymath}
\begin{displaymath}
+C_{01}(23.29-\frac{13.71}{r^2}-8.11r^2+0.066r^4)+C_{11}(-76.83+\frac{41.48}{r^2}+
62.06r^2-12.66r^4),
\end{displaymath}
\begin{equation}
A_2=-8.35+2.34r^2+C_{20}(8.35-2.34r^2)+C_{02}(1.80-2.34r^2)+
\end{equation}
\begin{displaymath}
+C_{10}(30.38-26.93r^2+4.00r^4)
+C_{01}(5.96-9.62r^2+2.69r^4)+C_{11}(-19.22+ 36.18r^2-23.80r^4),
\end{displaymath}
\begin{equation}
A_3=-0.99+0.99r^2+C_{20}(0.99-0.99r^2)+C_{02}(1.98-0.99r^2)+
\end{equation}
\begin{displaymath}
+C_{10}(1.65-9.72r^2+1.50r^4)
+C_{01}(0.97-2.21r^2+0.94r^4)+C_{11}(-1.62+ 4.02r^2-5.24r^4).
\end{displaymath}

\vspace{3mm}
{\underline {$e^++e^-\to J/\Psi+\chi_{c1}$}}

\begin{equation}
B_1=-0.002+2.66r^2+C_{20}(0.002-2.66r^2)-2.66r^2C_{02}+
\end{equation}
\begin{displaymath}
+C_{10}(-6.17-14.23r^2+4.59r^4)
+C_{01}(-4.62-7.14r^2+3.13r^4)+C_{11}(11.56+28.13r^2-27.18r^4),
\end{displaymath}
\begin{equation}
B_2=-2.10-\frac{0.004}{r^2}+C_{20}(2.10+\frac{0.004}{r^2})+C_{10}(14.99-
\frac{12.39}{r^2}-3.34r^2+0.016r^4)+
\end{equation}
\begin{displaymath}
+C_{01}(7.29-\frac{9.28}{r^2}-0.83r^2+0.01r^4)+C_{11}(-24.44+\frac{23.21}{r^2}+10.13r^2-0.98r^4),
\end{displaymath}
\begin{equation}
B_3=-1.63r^2+1.63r^2C_{20}+1.63r^2C_{02}+C_{10}(10.64r^2-2.87r^4)+
\end{equation}
\begin{displaymath}
+C_{01}(5.43r^2-1.20r^4)+C_{11}(-16.41r^2+13.44r^4),
\end{displaymath}
\begin{equation}
B_4=-0.002+0.82r^2+C_{20}(0.002-0.82r^2)-0.82r^2C_{02}+
\end{equation}
\begin{displaymath}
+C_{10}(-6.17-1.23r^2+1.77r^4)+C_{11}(11.56+1.35r^2-11.46r^4).
\end{displaymath}

\vspace{3mm}
{\underline {$e^++e^-\to J/\Psi+\chi_{c2}$}}

\begin{equation}
C_1=-1.84+1.84C_{20}+C_{10}(7.50-1.87r^2+0.01r^4)+
\end{equation}
\begin{displaymath}
+C_{01}(3.72-0.97r^2+0.02r^4)+C_{11}(-23.15+11.46r^2-0.95r^4),
\end{displaymath}
\begin{equation}
C_2=2.63r^2-2.63r^2C_{20}-2.63r^2C_{02}+C_{10}(-13.28r^2+4.55r^4)+
\end{equation}
\begin{displaymath}
+C_{01}(-7.72r^2+2.96r^4)+C_{11}(44.33r^2-30.95r^4),
\end{displaymath}
\begin{equation}
C_3=1.59r^2-1.59r^2C_{20}-1.59r^2C_{02}+C_{10}(-5.58r^2+2.85r^4)+
\end{equation}
\begin{displaymath}
+C_{01}(-3.91r^2+1.10r^4)+C_{11}(12.99r^2-14.29r^4),
\end{displaymath}
\begin{equation}
C_4=-0.80r^2-0.80r^2C_{20}+0.80r^2C_{02}+C_{10}(1.41r^2-1.75r^4)+
\end{equation}
\begin{displaymath}
+C_{01}(4.02r^2-1.51r^4)+C_{11}(-4.43r^2+12.19r^4),
\end{displaymath}
\begin{equation}
C_5=-2.39r^2-2.39r^2C_{20}+2.39r^2C_{02}+C_{10}(6.91r^2-4.33r^4)+
\end{equation}
\begin{displaymath}
+C_{01}(10.29r^2-2.98r^4)+C_{11}(-32.09r^2+32.60r^4),
\end{displaymath}
\begin{equation}
C_6=-2.37r^4C_{10}+C_{11}r^2(-12.59r^2+1.65r^4),
C_7=3.29r^4C_{10}+C_{11}r^2(-12.59r^2+1.65r^4).
\end{equation}

\vspace{3mm}
{\underline {$e^++e^-\to \eta_c+h_c$}}

\begin{equation}
D_1=-1.58+1.29r^2+C_{20}(1.58-1.29r^2)+C_{02}(1.58-1.29r^2)+
\end{equation}
\begin{displaymath}
+C_{10}(0.70-9.33r^2+2.97r^4)
+C_{01}(1.35-5.29r^2+1.01r^4)+C_{11}(-0.60+19.42r^2-16.51r^4),
\end{displaymath}
\begin{equation}
D_2=0.14-1.14r^2+C_{20}(-0.14+1.14r^2)+C_{02}(-3.73+1.14r^2)+
\end{equation}
\begin{displaymath}
+C_{10}(7.11+2.77r^2+2.37r^4)
+C_{01}(-0.12+4.72r^2-1.54r^4)+C_{11}(-6.08-3.69r^2-12.92r^4),
\end{displaymath}
\begin{equation}
D_3=-1.49+\frac{3.47}{r^2}+C_{20}(1.49-\frac{3.47}{r^2})+C_{10}(12.21-
\frac{15.99}{r^2}-2.35r^2+0.02r^4)+
\end{equation}
\begin{displaymath}
+C_{01}(6.76-\frac{2.97}{r^2}-0.80r^2+0.02r^4)+C_{11}(-32.42+
\frac{13.67}{r^2}+13.22r^2-1.35r^4).
\end{displaymath}

\section{The functions $F_i^{(k)}(r^2)$ (k=0,1,2,3) entering in the
production cross section (18)}

{\underline {$e^++e^-\to J/\Psi+\chi_{c0}$}}

\begin{equation}
F_0^{(0)}=12.94r^2+690.48r^4-391.86r^6+39.82r^8+6.36r^{10},
\end{equation}
\begin{equation}
F_1^{(0)}=-43.18r^2-4554.04r^4+5941.64r^6-1942.48r^8+89.61r^{10},
\end{equation}
\begin{equation}
F_2^{(0)}=-25.87r^2-1380.96r^4+783.71r^6-79.64r^8-12.72r^{10},
\end{equation}
\begin{equation}
F_3^{(0)}=36.04r^2+7634.68r^4-15478.8r^6+10011.2r^8-2095.47r^{10},
\end{equation}
\begin{equation}
F_4^{(0)}=-25.29r^2-996.50r^4+1961.67r^6-957.35r^8+97.10r^{10},
\end{equation}
\begin{equation}
F_5^{(1)}=-38.81r^2-447.77r^4+601.11r^6-101.74 r^8 -12.87r^{10},
\end{equation}
\begin{equation}
F_6^{(1)}=12.36r^2+349.41 r^4-1216.3 r^6 +1370.5 r^8 -579.05 r^{10},
\end{equation}
\begin{equation}
F_7^{(1)}=84.44r^2+6382.47r^4-15011.8r^6+13175.8r^8-4056.79 r^{10}.
\end{equation}

\vspace{1mm}
{\underline {$e^++e^-\to J/\Psi+\chi_{c1}$}}

\begin{equation}
F_0^{(1)}=165.06r^4-248.34r^6+75.88r^8+17.33r^{10},
\end{equation}
\begin{equation}
F_1^{(1)}=-1655.33r^4 + 3347.61r^6 -1829.98r^8 + 80.46r^{10},
\end{equation}
\begin{equation}
F_2^{(1)}=-330.12r^4 + 496.69r^6 -151.76r^8 - 34.67r^{10},
\end{equation}
\begin{equation}
F_3^{(1)}=4263.79r^4 - 10572.2r^6 +8306.62r^8-1953.72r^{10},
\end{equation}
\begin{equation}
F_4^{(1)}=-752.14r^4 + 1660.5r^6 -979.66r^8 + 60.00r^{10},
\end{equation}
\begin{equation}
F_5^{(1)}=-237.01r^4 + 390.21r^6 -117.85r^8 - 35.64r^{10},
\end{equation}
\begin{equation}
F_6^{(1)}=920.55 rr^4 - 2520.4r^6 +2230.78r^8 - 612.50r^{10},
\end{equation}
\begin{equation}
F_7^{(1)}=7012.32r^4 -18599.5r^6+15704.7r^8-4122.93r^{10}.
\end{equation}

\vspace{1mm}
{\underline {$e^++e^-\to J/\Psi+\chi_{c2}$}}

\begin{equation}
F_0^{(2)}=23.41r^2 +99.63r^4 -247.93r^6 +110.30r^8 +27.28r^{10},
\end{equation}
\begin{equation}
F_1^{(2)}=-78.15r^2 -931.82r^4 + 2499.9r^6 -1804.84r^8+137.34r^{10},
\end{equation}
\begin{equation}
F_2^{(2)}=-46.82r^2 -199.27r^4 + 495.86r^6 - 220.60r^8 -54.56r^{10},
\end{equation}
\begin{equation}
F_3^{(2)}=65.23r^2+1952.35r^4-5759.03r^6 + 5433.52r^8-1541.06r^{10},
\end{equation}
\begin{equation}
F_4^{(2)}=-94.75r^2-1014.54r^4 +2475.82r^6-1509.41r^8 +50.08r^{10},
\end{equation}
\begin{equation}
F_5^{(2)}=-70.23r^2-213.25r^4+608.04r^6-269.46r^8-55.66r^{10},
\end{equation}
\begin{equation}
F_6^{(2)}=95.89r^2+2191.11r^4-5632.06r^6+4350.3r^8-954.85 rr^{10},
\end{equation}
\begin{equation}
F_7^{(2)}=316.34r^2+7972.42r^4-22933.8r^6+20774.7r^8-5658.04r^{10}.
\end{equation}

\vspace{1mm}
{\underline {$e^++e^-\to \eta_c+h_c$}}

\begin{equation}
F_0^{(3)}=11.69r^2-27.29r^4 + 35.00r^6 - 22.48r^8 + 6.02r^{10},
\end{equation}
\begin{equation}
F_1^{(3)}=-10.35r^2+41.40r^4-185.00r^6+250.24r^8-141.63r^{10},
\end{equation}
\begin{equation}
F_2^{(3)}=-23.375r^2 +54.57r^4-70.00r^6 + 44.97r^8-12.04r^{10},
\end{equation}
\begin{equation}
F_3^{(3)}=2.29r^2+88.53r^4+5.76r^6-348.17r^8+451.27r^{10},
\end{equation}
\begin{equation}
F_4^{(3)}=-19.99r^2 + 118.20r^4-215.81r^6 + 199.84r^8-86.92r^{10},
\end{equation}
\begin{equation}
F_5^{(3)}=-35.06r^2+101.73r^4-115.65r^6+61.33r^8-12.52r^{10},
\end{equation}
\begin{equation}
F_6^{(3)}=8.55r^2-81.12r^4+268.40r^6-363.38r^8 + 234.48r^{10},
\end{equation}
\begin{equation}
F_7^{(3)}=17.70r^2-263.94r^4 +959.68r^6-1671.7r^8+1317.3r^{10}.
\end{equation}

\section{Effective relativistic Hamiltonian}

For the calculation of the relativistic corrections in the bound
state wave functions $\Psi_0^{\cal S}$, $\Psi_0^{\cal P}$, we
consider the Breit potential (23). It contains a number of terms
which should be transformed in order to use the program of numerical
solution of the Schr\"odinger equation \cite{FFS}. The
rationalization of the kinetic energy operator can be done in the
following form \cite{Lucha}:
\begin{equation}
T=2\sqrt{{\bf p}^2+m^2}=2\frac{{\bf p}^2+m^2}{\sqrt{{\bf
p}^2+m^2}}\approx \frac{{\bf p}^2}{\tilde m}+\frac{2m^2}{\tilde E},
\end{equation}
where $\tilde m$ is the effective mass of heavy quarks,
\begin{equation}
\tilde m=\frac{\tilde E}{2}=\sqrt{{\bf p}^2_{eff}+m^2}.
\end{equation}
${\bf p}^2_{eff}$ should be considered as a new parameter which
effectively accounts for relativistic corrections in (C1). Numerical
values of ${\bf p}^2_{eff}$ for S- and P-wave charmonium states
discussed in \cite{Bodwin2,Bodwin4,apm2005} are presented in Table
III. In the case of S-wave states it is necessary to transform the
$\delta$-like terms of the potential. For this aim, we use the known
smeared $\delta$-function of the Gaussian form \cite{delta}:
\begin{equation}
\tilde\delta({\bf r})=\frac{b^3}{\pi^{3/2}}e^{-b^2r^2}
\end{equation}
with the additional parameter $b$ which defines the hyperfine
splitting in the $(c\bar c)$ system. Since the numerical results are
practically not dependent on $b$ in the range of commonly used
values $1.5\div 2.2$, we take $b=1.5$ GeV. The second term in the
Breit potential (23), which also has to be transformed, takes the
form:
\begin{equation}
\Delta\tilde U=-\frac{2\alpha_s}{3m^2r}\left[{\bf
p}^2-\frac{d^2}{dr^2}\right].
\end{equation}
In order to replace it by the effective term containing the
power-like potentials, we use the approximate charmonium wave
functions which can be written for S- and P-wave states as
\begin{equation}
\Psi_0^{\cal
S}(r)=\frac{\beta^{3/2}}{\pi^{3/4}}e^{-\frac{1}{2}\beta^2r^2},~~\Psi_0^{\cal
P}(r)=\sqrt{\frac{8}{3}}\frac{\beta^{3/2}}{\pi^{3/4}}\beta
re^{-\frac{1}{2}\beta^2r^2}Y_{1m}(\theta,\phi).
\end{equation}

\begin{table}
\caption{The parameters of the effective relativistic Hamiltonian.}
\bigskip
\begin{ruledtabular}
\begin{tabular}{|c|c|c|c|c|c|c|c|c|}
Meson $(c\bar c)$&$n^{2S+1}L_J$ &${\bf p}^2_{eff}$,~$GeV^2$ &
$\tilde m$,~GeV& $E$,
~GeV &$\beta$,~GeV& b,~GeV& $M^{th}$,~GeV& $M^{exp}$,~GeV ,\cite{PDG} \\
\hline $J/\Psi$&$1^3S_1$  & 0.5 &0.85 & 0.087 & 0.75 & 1.5& 3.044& 3.097  \\
\hline $\eta_c$&$1^1S_0$ & 0.5 & 0.85& 0.087  & 0.75& 1.5& 2.989& 2.980 \\
\hline $\chi_{c0}$&$1^3P_0$ &0.6 & 0.87& 0.479 & 0.55& ---& 3.437& 3.415\\
\hline $\chi_{c1}$&$1^3P_1$  & 0.6 & 0.87 & 0.479  & 0.55& ---&
3.479&
3.511\\  \hline $\chi_{c2}$&$1^3P_2$  & 0.6& 0.87 & 0.479 & 0.55& ---& 3.520& 3.556 \\
\hline $h_c$&$1^1P_1$  &0.6& 0.87 & 0.479  & 0.55& ---& 3.486& 3.526\\
\hline
\end{tabular}
\end{ruledtabular}
\end{table}

The wave functions (C5) give a good approximation of the true quark
bound state wave functions in the region of nonrelativistic momenta.
Using (C5), we transform (C4) as follows:
\begin{equation}
\Delta \tilde U\to\Delta\tilde
U^{eff}=-\frac{2\alpha_s}{3m^2r}\left(mE-mB+\beta^2\right)-\frac{8\alpha_s^2}{9mr^2}+
\frac{2\alpha_s A}{3m}+\frac{2\alpha_s\beta^4}{3m^2}r,
\end{equation}
where $E$ is the bound state energy of quarks which can be obtained
from the Schr\"odinger equation with the Hamiltonian $H_0$. In order
to derive (C6) we changed the operator ${\bf p}^2$ by its
nonrelativistic expression: ${\bf
p}^2\Psi_0=m[E+\frac{4\alpha_s}{3r}-Ar-B]\Psi_0$. As a result of
such transformations the potential $\Delta U_2$ takes the following
form in the case of S-states:\\[1mm]

{\underline{$J/\Psi$-meson:}}

\begin{equation}
\Delta
U_2(r)=\frac{20\pi\alpha_s}{9m^2}\frac{b^3}{\pi^{3/2}}e^{-b^2r^2}-
\frac{2\alpha_s}{3m^2r}(mE-mB+\beta^2)-\frac{26\alpha_s^2}{9mr^2}+
\frac{2\alpha_s\beta^4}{3m^2}r+\frac{2\alpha_sA}{3m},
\end{equation}

{\underline{$\eta_c$-meson:}}

\begin{equation}
\Delta
U_2(r)=-\frac{4\pi\alpha_s}{3m^2}\frac{b^3}{\pi^{3/2}}e^{-b^2r^2}-
\frac{2\alpha_s}{3m^2r}(mE-mB+\beta^2)-\frac{26\alpha_s^2}{9mr^2}+
\frac{2\alpha_s\beta^4}{3m^2}r+\frac{2\alpha_sA}{3m}.
\end{equation}
A similar transformation of the Breit Hamiltonian can be done for
the P-wave states. In Table III we present the results of the
calculation of the charmonium mass spectrum and a comparison with
the existing experimental data. The obtained masses agree with the
experimental ones within an accuracy $1\div 2$ per cent. So, we can
suppose that the obtained effective Hamiltonian allows to account
relativistic corrections in the bound state wave functions with
sufficiently good accuracy.


\begin{thebibliography}{99}
\bibitem{Belle}K. Abe, et al., Phys. Rev. D {\bf 70}, 071102 (2004).
\bibitem{BaBar}B. Aubert, et al., Phys. Rev. D {\bf 72}, 031101 (2005).
\bibitem{BL1}E. Braaten, J. Lee, Phys. Rev. D {\bf 67}, 054007 (2003); Phys.
Rev. D {\bf 72}, 099901(E) (2005).
\bibitem{Chao}K.-Y. Liu, Z.-G. He, K.-T. Chao, Phys. Lett. B {\bf 557}, 45 (2003).
\bibitem{Qiao}K. Hagiwara, E. Kou, C.-F. Qiao, Phys. Lett. B {\bf
570}, 39 (2003).
\bibitem{BLB}G.T. Bodwin, J. Lee, E. Braaten, Phys. Rev. Lett. {\bf
90}, 162001 (2003).
\bibitem{BGL}S.J. Brodsky, A.S. Goldhaber, J.Lee, Phys. Rev. Lett.
{\bf 91}, 112001 (2003).
\bibitem{Chao1}K.-Y. Liu, Z.-G. He, K.-T. Chao, Phys. Rev. D {\bf 77}, 014002 (2008).
\bibitem{Ma}J.P. Ma, Z.G. Si, Phys. Rev. D {\bf 70}, 074007 (2004).
\bibitem{BC}A.E. Bondar, V.L. Chernyak, Phys. Lett. B {\bf 612}, 215 (2005).
\bibitem{BLL}V.V.Braguta, A.K. Likhoded, A.V. Luchinsky, Phys. Rev.
D {\bf 72}, 074019 (2005).
\bibitem{ZGC}Y.-J. Zhang, Y.-J. Gao, K.-T. Chao, Phys. Rev. Lett.
{\bf 96}, 092001 (2006).
\bibitem{Bodwin2}G.T. Bodwin, D. Kang, J. Lee, Phys.Rev. D {\bf 74}, 114028 (2006).
\bibitem{EM2006}D. Ebert, A.P. Martynenko, Phys. Rev. D {\bf 74}, 054008 (2006).
\bibitem{Ji}H.-M.Choi, Ch.-R. Ji, Phys. Rev. D {\bf 76}, 094010, (2007).
\bibitem{He}Z.-G. He, Y.Fan, K.-T. Chao, Phys. Rev. D {\bf 75}, 074011 (2007).
\bibitem{AVB}A.V. Berezhnoy, Phys. Atom. Nucl. {\bf 71}, 1803 (2007).
\bibitem{Bodwin4}G.T. Bodwin, J. Lee, Ch.Yu, Phys. Rev. {\bf D77}, 094018 (2008).
\bibitem{EFGM2009}D. Ebert, R.N. Faustov, V.O. Galkin, A.P.
Martynenko, Phys. Lett. {\bf B672}, 264 (2009).
\bibitem{apm2005}A.P. Martynenko, Phys. Rev. D {\bf 72}, 074022 (2005).
\bibitem{rqm5}D. Ebert, R.N. Faustov, V.O. Galkin, A.P. Martynenko,
Phys. Rev. D  {\bf 70}, 014018 (2004).
\bibitem{F1973}R.N. Faustov, Ann. Phys. {\bf 78}, 176 (1973).
\bibitem{BP}S.J. Brodsky, J.R. Primack, Ann. Phys. {\bf 52}, 315
(1969).
\bibitem{Kuhn}J.H. Kuhn, J. Kaplan, El J. O. Safiani, Nucl. Phys.
{\bf B157}, 125 (1979).
\bibitem{FORM}J.A.M. Vermaseren, FORM, e-preprint math-ph/0010025.
\bibitem{BBL}G.T. Bodwin, E. Braaten, G.P. Lepage, Phys. Rev. {\bf
D51}, 1125 (1995).
\bibitem{pot1}N. Brambilla, A. Pineda, J. Soto, A. Vairo, Rev. Mod.
Phys. {\bf 77}, 1423 (2005).
\bibitem{pot2}B. Kniehl, A.A. Penin, V.A. Smirnov, M. Steinhauser,
Nucl. Phys. {\bf B635}, 357 (2002).
\bibitem{pot3}K. Melnikov, A. Yelkhovsky, Phys. Rev. {\bf D59},
114009 (1999).
\bibitem{FFS}P. Falkensteiner, H. Grosse, F.F. Sch\"oberl, P. Hertel,
Comp. Phys. Comm. {\bf 34}, 287 (1985).
\bibitem{BLL1}V.V. Braguta, A.K. Likhoded, A.V. Luchinsky, Phys.
Lett. {\bf B635}, 299 (2006).
\bibitem{Chao2008}Y.-J. Zhang, Y.-Q. Ma, K.-T. Chao, Phys. Rev.
{\bf D78}, 054006 (2008).
\bibitem{PDG}Particle Data Group, J. Phys. G {\bf 33}, 1 (2006).
\bibitem{rqm1}D. Ebert, R.N. Faustov, V.O. Galkin, Phys. Rev. {\bf
D67}, 014027 (2002).
\bibitem{QWG}N. Brambilla et al. Heavy Quarkonium Physics, FERMILAB
Report, Report No. FERMILAB-FN-0779, CERN Yellow Report, Report No.
CERN-2005-005.
\bibitem{Lucha}W. Lucha, F.F. Sch\"oberl, M. Moser, Preprint
HEPHY-PUB 594/93.
\bibitem{delta}I.M. Narodetskii, Yu.A. Simonov, V.P. Yurov, Yad.
Fiz. {\bf 55}, 2818 (1992).
\end{thebibliography}
\end{document}